\documentclass[12pt,a4paper,english]{article}
\usepackage[dvips]{graphicx}
\usepackage{epsfig}
\usepackage{float}
\usepackage{times}
\usepackage{hhline}
\usepackage[english]{babel}
\usepackage[latin1]{inputenc}
\usepackage[T1]{fontenc}
\usepackage{amssymb}
\usepackage{amsmath}
\usepackage{lscape}
\usepackage{tabularx}
\numberwithin{equation}{section}
\begin{document}

% ------------ ETUSIVU --------------------------------------

\thispagestyle{empty}

\begin{center}

\vspace*{3.5cm}

\vspace{1.0cm} {\Huge Lemaitre-Tolman-Bondi model and accelerating expansion}

\vspace{1.0cm} Authors: Kari Enqvist

\vspace{1.0cm}
Department of Physical Sciences and Helsinki Institute of Physics,\\
FIN-00014 University of Helsinki

\vspace{1.0cm} \today

\end{center}

\vspace{2.0 cm}

%--------TIIVISTELMÄ----------
\begin{abstract}

I discuss the spherically symmetric but inhomogeneous Lemaitre-Tolman-Bondi
(LTB) metric, which provides an exact toy model for an inhomogeneous universe.
Since we observe light rays from the past light cone, not the expansion
of the universe, spatial variation in matter density and  Hubble rate
can have the same effect on redshift as acceleration in a perfectly
homogeneous universe. As a consequence, a simple spatial variation in the Hubble rate can account for
the distant supernova data in a dust universe without any dark energy.
I also review various attempts towards a semirealistic description
of the universe based on the LTB model.

\end{abstract}

\vspace{1.0cm}


%------------- TEXT -----------------------------------

\newpage
\pagenumbering{arabic}

\section{Introduction}

The simplest homogeneous and isotropic cosmological models, based on
the Friedmann-Robertson-Walker (FRW) metric, have proved to be remarkably
successful ever since Edwin Hubble in 1929 rather cautiously suggested that
the apparent linear correlation between the observed redshifts and distances of 24 galaxies
could hint towards the possibility "that the velocity-distance
relation may represent the de Sitter effect, and hence that numerical
data may be introduced into discussions of the general curvature of
space" \cite{hubble1929}. Indeed, numerical data now guides the
development of cosmology, which has become a precision science,
albeit mostly within the framework of a perfectly homogenous background metric.

The FRW universe is characterized by two
functions, the Hubble rate $H$ and the density parameter
$\Omega$, or the average expansion rate
and the average density of mass energy, respectively, which depend on time but are independent
of the spatial location. However, one
should keep in mind that their values cannot be extracted directly
from the observations but must be deduced
from the properties of light coming from the past light cone. In the context of the FRW model
this is almost trivial, since the redshift $z$ and scale factor $a(t)$ are
everywhere related by $z=a(t_o)/a(t_e)-1$, where the subscripts refer respectively
to the observation and the emission of light. This theoretical simplicity should however
not cloud the fact that all cosmological parameter determination requires  an element of interpretation of the data.
Of course, the FRW interpretation of
the properties of the past light cone has served cosmology well, giving a good fit to
observations and, until the late 90's,
implying a matter dominated universe with $\Omega\approx \Omega_M$.

The situation changed dramatically with the WMAP \cite{WMAP} and distant supernova
 data \cite{SNs}. Considering the recent data from
supernovae \cite{Riess:2004nr, Astier:2005qq}, galaxy
distributions \cite{Eisenstein:2005su} and anisotropies of the
cosmic microwave background \cite{Spergel:2006hy}, the simplest FRW
model would now lead to a highly contradictory picture of the universe,
with the following best fit values for the average matter density:
\begin{itemize}
\item Cosmic microwave background: $\Omega_M \sim 1$
\item Galaxy surveys: $\Omega_M \sim 0.3$
\item Type Ia supernovae: $\Omega_M \sim 0 $
\end{itemize}
As is well known, the
glaring discrepancies between the different data sets have conventionally been
remedied by introducing the cosmological constant $\Lambda$
or vacuum energy $\Omega_\Lambda$ to the Einstein equations. This gives rise to an accelerated expansion of the universe.
As a consequence, the apparent dimming of the luminosity of distant supernovae finds,
in the context of perfectly homogeneous universe, a natural explanation\footnote{Even
if the primordial perturbation is not scale free, the combination of the CMB
fluctuations and the shape of the correlation function up to $\sim 100 h^{-1}$Mpc,
seems to require dark energy for a homogeneous FRW model \cite{subiretal}.}.

However, although the cosmological concordance $\Lambda$CDM-model \cite{Copeland:2006wr}
fits the observations well, there is no theoretical understanding of the origin of the
cosmological constant or its magnitude. For particles physicists, who have spent a long time trying
to prove that the cosmological constant must be zero, the tremendously small cosmological constant
which just now happens to start to dominate the energy budget of the universe,
is a theoretical nightmare. There exist a large number
of different dark energy models (see e.g. \cite{Copeland:2006wr, Straumann:2006tv}) that attempt to
provide a dynamical explanation for the cosmological constant, but none of them are
compelling from particle physics point of view; moreover, very often they
require fine-tuning. Modifications of the
general theory of relativity on cosmological scales appear to suffer
from analogous problems. For instance, $f(R)$ gravity theories \cite{fR} in the
metric formalism are plagued by instabilities \cite{dolgovkawasaki} while in the Palatini approach
the cosmological constant seems to be essentially the only consistent
modification that fits all the cosmological data
\cite{Koivisto:2006ie}.

Facing such difficulties, one might be tempted to consider relinquishing the
FRW assumption of the
perfect homogeneity of the universe. After all, inhomogeneities are abundant in
the universe: there are not only clusters of galaxies but also large voids.
Because general relativity is a non-linear theory, even relatively small local
inhomogeneities with a sufficiently large density contrast could in principle give rise to
cosmological evolution that is not accessed by the usual
cosmological perturbation theory in an FRW background. In fact,
the potentially interesting consequences of the inhomogeneities were recognized
already at the time when the homogeneous and isotropic
models of the universe were first studied, but their impact on the
global dynamics of the universe is still largely unknown (see e.g.\
\cite{Krasinski}). Then the question arises: could the acceleration of the
universe be just a trick of light, a misinterpretation that arises due to
the oversimplification of the real, inhomogeneous universe inherent in the FRW model?
Light, while traveling though inhomogeneities, does not
see the average Hubble expansion but rather feels its variations, which could sum up to an
important correction\footnote{For a recent calculation of the small scale inhomogeneity-induced
correction to the cosmological constant that one would infer from
an analysis of the luminosities and redshifts of Type Ia supernovae, assuming a homogeneous universe, see \cite{systcorr}.}.
This effect is particularly important for the case of large scale inhomogeneities which will be the focus
of the present paper. If the local Hubble expansion rate were to vary smoothly at scales of
the order of, say, thousand megaparsecs, that would very much change our interpretation of
the distant supernova redshifts.
In such an inhomogeneous universe we could also just happen to be located in a special position. For instance,
fate could have relegated us to an
underdense region with a larger than average local Hubble parameter so that
the discrepancy between nearby and distant supernovae luminosities could
be resolved without dark energy.

Local inhomogeneities have recently been invoked as the culprit for the
apparent acceleration of the expansion of the
universe\footnote{Inhomogeneities as an alternative to dark energy were first discussed in
\cite{Pascual-Sanchez:1999zr}.}, in particular
by virtue of their so-called backreaction on the metric (for a discussion on the issues
involved and a comprehensive list of references, see
\cite{Rasanen:2006kp,buchert:2007}). One constructs an
effective description of the universe by averaging out the inhomogeneities
to obtain averaged, effective Einstein equations which, in
addition to the terms found in the usual homogeneous case, include new terms that represent the
effect of the inhomogeneities \cite{Buchert:1999er,Ellis:2005uz,Coley:2005ei}.

However, since we can only observe the redshift and energy
flux of light arriving from a given source, not the expansion rate or the
matter density of the universe nor their averages, one may wonder what are the
actual observables related to the averaged equations. To wit, since
we do not observe the average expansion of the universe
directly, its average acceleration is also an indirect conclusion, arising
from the fact that \textit{in the perfectly homogeneous cosmological
models} dark energy is required for a good fit. Consequently, there is
no a priori reason to assume that an accelerated expansion is necessarily required
to fit the data if one assumes a
general inhomogeneous model of the universe. One may
also add that the averaging procedure as such is not without problems:
in general it is not correct to
integrate out constrained degrees of freedom as if they were
independent, and in cosmology the fact that we can make observations only along
our past light cone makes the observable universe a constrained
system. Hence it would be desirable to study the effects of the
the inhomogeneities on the directly observable
light in an exact cosmological model. Unfortunately,  in the presence of generic
inhomogeneities this would be practically an impossible task. Instead, one
must resort to toy models, the simplest of which is the spherically symmetric but inhomogeneous
Lemaitre-Tolman-Bondi (LTB) model \cite{Lemaitre:1933qe,Tolman:1934za,Bondi:1947av}.

The great virtue of the LTB model is that it is exact.
Because of its high degree of symmetry, it may not be realistic as such,
but the LTB model is nevertheless interesting at least on
two counts. First, it serves as a simple testing ground for the effects of inhomogeneities
when fitting the cosmological data without dark energy. Second, since the fits
can be performed unambiguously, the nature of
the effective accelaration in the models where the spatial
degrees of freedom have been averaged out,
can be made transparent by comparing the averaged and "exact" models.

Of course, one can also take the LTB model more seriously.
For instance,
one may use the LTB metric to describe
a local underdense bubble in FRW universe, for which there is some evidence both from supernova \cite{zehavi98}
and galaxy data \cite{localhole}.
First attempts along these directions \cite{Tomita} assumed an underdense region separated from
the outside homogeneous FRW universe by a singular mass shell, followed by investigations
of
more realistic models with a continuous transition between the inner underdensity and
the outer homogeneous universe (see e.g. \cite{Alnes:2005rw, moffat}).
More complicated situations, including off-centered observers, can also be
addressed, as will be discussed in Sect. \ref{realistic}.

\section{The Lemaitre-Tolman-Bondi metric}\label{ltbmetric}

Let us consider a spherically symmetric dust universe with radial
inhomogeneities as seen from our location at the center. Choosing
spatial coordinates to comove ($dx^i/dt = 0$) with the matter, the
spatial origin ($x^i=0$) as the symmetry center, and the time
coordinate ($x^0 \equiv t$) to measure the proper time of the
comoving fluid, the line element takes the general form \cite{Lemaitre:1933qe,Tolman:1934za,Bondi:1947av}
\begin{equation}\label{metric1}
ds^2 = - dt^2 + X^{2}(r,t)dr^2 + A^{2}(r,t) \left( d\theta^2 +
\sin^2 \theta d\varphi^2 \right)~,
\end{equation}
where the functions $A(r,t)$ and $X(r,t)$ have
both temporal and spatial dependence. The
homogeneous FRW-metric is a special case and is obtained by letting
\begin{equation}
X (r,t) \rightarrow
\frac{a(t)}{\sqrt{1-kr^2}},~~ A(r,t) \rightarrow a(t)r.
\end{equation}
The energy momentum tensor is given
by
\begin{equation}\label{energy}
T^{\mu}_{\phantom{\mu} \nu} = - \rho_M (r,t) \delta^{\mu}_{0}
\delta^{0}_{\nu} - \rho_\Lambda \delta^{\mu}_{\phantom{\mu} \nu}~,
\end{equation}
where $\rho_M(r,t)$ is the matter density, $u^\mu =
\delta^{\mu}_{0}$ represent the components of the 4-velocity-field
of the fluid, and we have kept the vacuum energy $\rho_\Lambda$ for
generality. Note that although the fluid is staying at fixed spatial
coordinates, it can physically move in the radial direction.
Plugging Eq. (\ref{metric1}) into the Einstein equation,
$G^{\mu}_{\phantom{\mu} \nu} = 8 \pi G T^{\mu}_{\phantom{\mu} \nu}$,
one finds the set of equations
\begin{equation}\label{yht00'}
-2\frac{A''}{AX^2}+2\frac{A'X'}{AX^3}+2\frac{\dot{X}\dot{A}}{AX}+\frac{1}{A^2}+\left(\frac{\dot{A}}{A}\right)^2
-\left(\frac{A'}{AX}\right)^2 = 8 \pi G (\rho_M + \rho_\Lambda)~,
\end{equation}

\begin{equation}\label{yht10'}
\dot{A}'=A'\frac{\dot{X}}{X}~,
\end{equation}

\begin{equation}\label{yht11'}
2\frac{\ddot{A}}{A}+\frac{1}{A^2}+\left(\frac{\dot{A}}{A}\right)^2-\left(\frac{A'}{AX}\right)^2
= 8 \pi G \rho_\Lambda~,
\end{equation}
and
\begin{equation}\label{yht22'}
-\frac{A''}{AX^2}+\frac{\ddot{A}}{A}+\frac{\dot{A}}{A}\frac{\dot{X}}{X}+\frac{A'X'}{AX^3}+\frac{\ddot{X}}{X}
= 8 \pi G \rho_\Lambda~.
\end{equation}
These contain only three independent
differential equations, and we may solve $\dot{X}$ and $\ddot{X}$ from
Eq. (\ref{yht10'}) and $A'^2$ and $A''$ from Eq.
(\ref{yht11'}). Then one can substitute these into Eq.
(\ref{yht22'}) and find that it yields an identity. Thus only two of
equations (\ref{yht10'})-(\ref{yht22'}) are independent. One can
easily solve Eq. (\ref{yht10'}) to obtain
\begin{equation}\label{äx}
X(r,t)=C(r)A'(r,t)~,
\end{equation}
where the function $C(r)$ depends only on the coordinate $r$. By
redefining $C(r)\equiv {1}/{\sqrt{1-k(r)}}$, where $k(r)<1$, we
can thus write the LTB metric Eq. (\ref{metric1}) in its usual form:
\begin{equation}\label{metric}
ds^2 = - dt^2 + \frac{(A'(r,t))^2}{1-k(r)}dr^2 + A^{2}(r,t) \left(
d\theta^2 + \sin^2 \theta d\varphi^2 \right)~,
\end{equation}
where $k(r)$ is a function associated with the curvature of
$t={\rm{const.}}$ hypersurfaces. The FRW metric is the limit $A(r,t) \rightarrow a(t)r$ and $ k(r)
\rightarrow k r^2$.

The two independent equations are given by
\begin{eqnarray}
\label{yht00}
\frac{\dot{A}^2+k(r)}{A^2} + \frac{2 \dot{A} \dot{A}' + k'(r)}{A A'}
&=& 8 \pi G (\rho_M + \rho_\Lambda),\\
\label{yht11}
\dot{A}^2 + 2 A \ddot{A} + k(r) &=& 8 \pi G \rho_\Lambda A^2~.
\end{eqnarray}
The first integral of Eq. (\ref{yht11}) is
\begin{equation}\label{int11}
\frac{\dot{A}^2}{A^2} = \frac{F(r)}{A^3} + \frac{8 \pi G}{3}
\rho_\Lambda - \frac{k(r)}{A^2}~,
\end{equation}
where $F(r)$ is a non-negative function that, like $k(r)$, is fixed by the
boundary condition. Substituting Eq.
(\ref{int11}) into Eq. (\ref{yht00}) yields
\begin{equation}\label{yht0011}
\frac{F'}{A'A^2} = 8 \pi G \rho_M~.
\end{equation}
By combining Eqs. (\ref{yht00}) and (\ref{yht11}) we can construct
the generalized acceleration equation
\begin{equation}\label{2ndfriedmann}
\frac{2}{3} \frac{\ddot{A}}{A} + \frac{1}{3} \frac{\ddot{A}'}{A'} =
- \frac{4 \pi G}{3} (\rho_M - 2 \rho_\Lambda)~
\end{equation}
which implies that the total acceleration, represented by the
left hand side, is negative everywhere unless the vacuum energy is
large enough: $\rho_\Lambda > \rho_M/2$. However, it does not
exclude the possibility of having radial acceleration
($\ddot{A}'(r,t)>0$), even in the pure dust universe, if the angular
scale factor $A(r,t)$ is decelerating fast enough, and vice versa. This
serves to demonstrate how the very notion of the
acceleration becomes ambiguous in the presence of the
inhomogeneities \cite{Apostolopoulos:2006eg}.

The boundary condition functions $F(r)$ and $k(r)$ are specified by
the exact physical nature of the inhomogeneities. Their relation to the FRW
model parameters can be recognized by comparing Eq.
(\ref{int11}) with the Einstein equation for the homogeneous FRW-model
\begin{eqnarray}
\label{frweq}
H^2(t) &\equiv &\frac{\dot{a}(t)}{a(t)} = \frac{8 \pi
G}{3}(\rho_{M}+\rho_{\Lambda})-\frac{k}{a^2}\\
&=& H_0^2 \left[\Omega_M \left( \frac{a_0}{a} \right)^3 +
\Omega_\Lambda + (1- \Omega_\Lambda - \Omega_M ) \left(
\frac{a_0}{a} \right)^2 \right]~,
\end{eqnarray}
where $a_0 \equiv a(t_0)$ and $H_{0} \equiv H(t_{0})$. Thus, a
comparison between Eqs. (\ref{int11}) and (\ref{frweq}) motivates one to
define the local Hubble rate as
\begin{equation}\label{hupple}
H(r,t) \equiv \frac{\dot{A}(r,t)}{A(r,t)}~.
\end{equation}
The local matter density can be defined through
\begin{equation}\label{isof}
F(r) \equiv H_0^2(r) \Omega_M(r) A_0^3(r)~,
\end{equation}
with
\begin{equation}\label{lillaf}
k(r) \equiv H_0^2(r) (\Omega_M(r) + \Omega_\Lambda(r) - 1)
A_0^2(r)~,
\end{equation}
where we have defined the boundary values at $t_0$
through $A_0(r) \equiv A(r,t_0)$, $H_0(r) \equiv H(r,t_0)$, and
$\Omega_\Lambda (r) \equiv {8 \pi G \rho_\Lambda}/{3 H_0^2(r)}$.
With these definitions, the position-dependent Hubble rate, Eq. (\ref{int11}), takes a physically
transparent form \cite{teppo}:
\begin{equation}\label{Friidman}
H^2(r,t) = H_0^2(r) \left[ \Omega_M(r) \left(\frac{A_0}{A} \right)^3
+ \Omega_\Lambda (r) + \Omega_c(r) \left(\frac{A_0}{A} \right)^2
\right]~,
\end{equation}
where $\Omega_c(r) \equiv 1- \Omega_\Lambda (r)-\Omega_M(r)$.

The
difference between the conventional Friedmann equation (\ref{frweq})
and its LTB generalization, Eq. (\ref{Friidman}), is that all the
quantities in the LTB case depend on the $r$-coordinate.
Thus in the presence of inhomogeneities, the values of the
Hubble rate and the matter density can vary at every spatial point so that
the inhomogeneous dust models are defined by two functions of the
spatial coordinates: $H_0(x^i)$ and $\Omega_M(x^i)$. As a
consequence, the inhomogeneities are of two physically different
kinds: inhomogeneities in the matter distribution, and
inhomogeneities in the expansion rate. Although their dynamics are
coupled via the Einstein equation, as boundary conditions they are
independent. The universe could have an inhomogeneous big bang, where the
universe came into being at different times at different points, and/or
an inhomogeneous matter density. This opens up the possibility for an inhomogeneous universe that has
a homogeneous present-day $\Omega_M$; a model of this kind could potentially fit the
supernova data as well as the galaxy surveys without invoking dark
energy. However, if  $\Omega_M(r) = {\rm{const.}}$, the physical matter
distribution $\rho_M$ itself has a spatial dependence provided
$H_0(r) \neq {\rm{const.}}$. It can be made constant by choosing $\Omega_M(r)H_0^2(r)={\rm const.}$

The spatial dependence holds
true even for the gauge freedom of the scale function. In the FRW
case the present value of the scale factor $a(t_0)$ can be chosen to
be any positive number. Similarly, the corresponding present-day
scale function $A(r,t_0)$ of the LTB model can be chosen to be any
smooth and invertible positive function. In what follows we will choose the
conventional gauge
\begin{equation}\label{gauge}
A(r,t_0)=r~.
\end{equation}
Integrating Eq. (\ref{Friidman}) then gives the relation between the scale factor $A(r,t)$ and the coordinates
$r$ and $t$, which can also be used to find the age of the LTB universe. One finds
\begin{equation}\label{secondintegral}
t_0-t = \frac{1}{H_0(r)}\int_{\frac{A(r,t)}{A_0(r)}}^{1} \frac{ dx}{
\sqrt{\Omega_M(r)  x^{-1} + \Omega_{\Lambda}(r)x^2+\Omega_{c}(r)}}~.
\end{equation}
For any space-time point with coordinates ($t,r,\theta,\varphi$),
Eq. (\ref{secondintegral}) determines the function $A(r,t)$ and all
its derivatives. Thus the metric Eq. (\ref{metric}) is specified,
and given the inhomogeneities, all the observable quantities can be
computed. Eq. (\ref{secondintegral}) can be integrated in terms of elementary
functions when $\Omega_{\Lambda}(r)=0$ or
$\Omega_{\Lambda}(r)+\Omega_{M}(r)=1$; as an example, in the latter case
one finds
\begin{equation}\label{A-tap4}
(t-t_0)H_0 = \frac{2}{3\sqrt{1-\Omega_M(r)}} \left[
{\rm{arsinh}}\sqrt{\omega(r) \left(
\frac{A(r,t)}{A_0(r)} \right)^3} -
{\rm{arsinh}}\sqrt{\omega(r)} \right]~,
\end{equation}
where
\begin{equation}
\omega(r)=\frac{1-\Omega_M(r)}{\Omega_M(r)}~.
\end{equation}
In this particular case $A(r,t)$ can be found explicitly as
\begin{equation}\label{Aomega1}
A(r,t)=A_0(r)\left[{\rm cosh}(\tau)+\sqrt{\frac{3}{8\pi G\rho_\Lambda}}H_0(r){\rm sinh}(\tau) \right]~,
\end{equation}
where $\tau=\sqrt{6\pi G\rho_\Lambda}(t-t_0)$.
\section{Inhomogeneities and luminosity distance}
\label{nullgeodesics}

To compare the inhomogeneous LTB model e.g. with the supernova
observations, we need an equation that relates the redshift and
energy flux of light with the exact nature of the inhomogeneities.
For this, one must study propagation  of light in the
LTB universe\footnote{Luminosity distance in a perturbed FRW universe
has been considered in \cite{BonvinDurrerGasparini}.}. Let us
here derive the appropriate equations for notational
clarity; a more general derivation for an off-center observer can be
found in \cite{Alnes:2006pf}.

From the symmetry of the situation, it is clear that light can
travel radially, that is, there exist geodesics with $d\theta=
d\varphi=0$. Moreover, since light always travels along null
geodesics, we have $ds^2=0$. Inserting these conditions into the
equation for the line element, Eq. (\ref{metric}), we obtain the
constraint equation for light rays
\begin{equation}\label{lightprop1}
\frac{dt}{du} = - \frac{dr}{du} \frac{A'(r,t)}{\sqrt{1-k(r)}}~,
\end{equation}
where $u$ is a curve parameter, and the minus sign indicates that we
are studying radially \textit{incoming} light rays.

Consider two light rays with solutions to Eq. (\ref{lightprop1})
given by $t_1 = t(u)$ and $t_2 = t(u) + \lambda(u)$. Inserting these
to Eq. (\ref{lightprop1}) we obtain
\begin{equation}\label{lightprop2}
 \frac{d}{du} t_1 = \frac{d t(u)}{du} = -
 \frac{dr}{du} \frac{A'(r,t)}{\sqrt{1-k(r)}}
\end{equation}
\begin{equation}\label{lightprop3}
\frac{d}{du} t_2 = \frac{dt(u)}{du} + \frac{d \lambda (u)}{du} = -
 \frac{dr}{du} \frac{A'(r,t)}{\sqrt{1-k(r)}} + \frac{d \lambda (u)}{du}
\end{equation}
\begin{equation}\label{lightprop4}
\frac{d}{du} t_2 = - \frac{dr}{du}
\frac{A'(r,t(u)+\lambda(u))}{\sqrt{1-k(r)}} = - \frac{dr}{du}
\frac{A'(r,t) + \dot{A}'(r,t) \lambda(u) }{\sqrt{1-k(r)}}~,
\end{equation}
where Taylor expansion has been used in the last step and only terms
linear in $\lambda(u)$ have been kept. Combining the right hand
sides of Eqs. (\ref{lightprop3}) and (\ref{lightprop4}) gives the
equality
\begin{equation}\label{lightprop5}
\frac{d \lambda (u)}{du} = - \frac{dr}{du} \frac{\dot{A}'(r,t)
\lambda(u)}{\sqrt{1-k(r)}}~.
\end{equation}
Differentiating the definition of the redshift, $z \equiv
({\lambda(0)-\lambda(u)})/{\lambda(u)}$, we obtain
\begin{equation}\label{lightprop6}
\frac{dz}{du} = - \frac{d \lambda(u)}{du} \frac{\lambda
(0)}{\lambda^2(u)} = \frac{dr}{du} \frac{(1+z) \dot{A}'
(r,t)}{\sqrt{1-k(r)}}~,
\end{equation}
where in the last step we have used Eq. (\ref{lightprop5}) and the
definition of the redshift. Finally, we can combine Eqs.
(\ref{lillaf}), (\ref{lightprop1}) and (\ref{lightprop6}) to obtain
the pair of differential equations
\begin{eqnarray}\label{dtdz}
\frac{dt}{dz} &=& \frac{-A'(r,t)}{(1+z) \dot{A}'(r,t)}~,\\
\label{drdz}
\frac{dr}{dz} &=&
\frac{\sqrt{1+H^2_0(r)(1-\Omega_M(r)-\Omega_\Lambda(r))A^2_0(r)}}{(1+z)
\dot{A}'(r,t)}~,
\end{eqnarray}
which determine the relations between the coordinates and the observable
redshift, i.e. $t(z)$ and $r(z)$.

Now that we have related the redshift to the inhomogeneities, we
still need the relation between the redshift and the energy flux
$F$, or the luminosity-distance, defined as $d_L \equiv \sqrt{{L}/{4
\pi F}}$, where $L$ is the total power radiated by the source. This is given by \cite{Ellis}
\begin{equation}\label{lumdist}
d_L (z) = (1+z)^2 A(r(z),t(z))~.
\end{equation}
Likewise,
the angular distance diameter is given by
\begin{equation}\label{angdist}
d_A (z) = A(r(z),t(z))~.
\end{equation}
As the $z$-dependence of $t$ and $r$ are determined by Eqs.
(\ref{dtdz}) and (\ref{drdz}) and the scale function $A(r,t)$ by Eq.
(\ref{secondintegral}), using Eq. (\ref{lumdist}) one can calculate
$d_L$ for a given $z$. All of these relations have a manifest
dependence on the inhomogeneities (i.e.\ on the functions $H_0(r)$
and $\Omega_M(r)$). What remains is a comparison of Eq.
(\ref{lumdist}) with the observed $d_L(z)$.

Because the boundary functions of the LTB model are arbitrary, it comes
as no surprise that any isotropic set of
observations can be explained by the appropriate inhomogeneities of the
LTB model \cite{Mustapha:1998jb}.
That the supernova data could be interpreted in terms of an inhomogeneous
LTB model with no cosmological constant was first suggested by Célérier \cite{celerier},
who pointed out that the LTB model is degenerate with respect to any magnitude-redshift relation so
that the accelerated expansion could be modelled by a very large number of inhomogeneity profiles.
In this sense the LTB model is not
predictive. The intriguing aspect here is rather the matter of principle which
the LTB model can be used to demonstrate: that the supernova data does
not necessarily imply accelerating expansion and hence the existence of dark
energy is not an unavoidable consequence of the data but rather depends on the
framework the data is interpreted in. Moreover, the inhomogeneities need not contradict
the observed homogeneity in galaxy surveys \cite{Eisenstein:2005su}, as is often claimed (see e.g. \cite{Vanderveld:2006rb}),
since the model admits
solutions with constant $\Omega_M$ but with a position-dependent $H$.

To demonstrate this, let us consider the gold sample of 157 supernovae
of Riess et.\ al.\
\cite{Riess:2004nr} and disregard LSS and CMB data for the moment.
In the FRW model the parameters that best describe our universe are
found by maximizing the likelihood function exp$({-\chi^2(
H_0,\Omega_M,\Omega_\Lambda))}$ constructed from the observations.
However, to find the boundary conditions of the LTB universe that
best describe our universe, we should in principle maximize the
likelihood \textit{functional} exp$({-\chi^2  (H_0 (r) , \Omega_M (r)
))}$. In practice, this is impossible. One can only consider
some physically motivated types for the functions $H_0 (r)$ and
$\Omega_M (r)$ that contain free parameters; these are then fitted
to the supernova observations by maximizing the leftover likelihood
function. In the literature there exist several  fits to the supernova data employing a simple LTB model with
different authors having chosen different density profiles (and, unfortunately, often
a different notation) \cite{Iguchi:2001sq,Alnes:2005rw,Garfinkle:2006sb, Chung:2006xh, teppo, Bolejko:2005fp}.

Since the expansion rate of the FRW universe has to accelerate in
order to fit the supernova data, the second time derivative of the
FRW scale function should be positive. In contrast, in the LTB universe the
observations are affected by the variation of all the dynamical
quantities along the past light cone, not just the time variation.
Indeed, the directional derivative along the past light cone is
given by
\begin{equation}\label{directionalderivative}
\frac{d}{dt} = \frac{\partial}{\partial t} +\frac{dt}{dr}
\frac{\partial}{\partial r} = \frac{\partial}{\partial t} -
\frac{A'(r,t)}{\sqrt{1-k(r)}} \frac{\partial}{\partial r}  \approx
\frac{\partial}{\partial t} - \frac{\partial}{\partial r},
\end{equation}
where the approximation in the last step is more accurate for
small values of $r$, but is qualitatively correct even for larger
$r$.

The main message of Eq. (\ref{directionalderivative}) is that from
the observational point of view, the negative $r$-derivative roughly
corresponds to the positive time derivative.  This is natural since
by looking at a source, we simultaneously look into the past (i.e.
along the \textit{negative} $t$-axis) and into a spatial distance (i.e.
along the \textit{positive} $r$-axis). Hence, to mimic the acceleration,
i.e.\ for the expansion rate to look as if it were to increase towards
us along the past light cone, the expansion $H_0(r)$ must
\textit{decrease} as $r$ grows: hence we should look for an LTB model with $H_0'(r)<0$.
Thus, keeping in mind the homogeneity of galaxy distributions, we could choose a simple four parameter LTB model like \cite{teppo}
\begin{eqnarray}\label{SNmodel}
H_0(r) &=& H + \Delta H e^{-r/r_0}~,\nonumber\\
\Omega_M(r) &=& \Omega_0 = {\rm{constant}}~,
\end{eqnarray}
where $H$,
$\Delta H$, $\Omega_0$ and $r_0$ are free parameters determined by the supernova
observations. The best fit values are found to be \cite{teppo}
\begin{equation}\label{SNfit}
H+ \Delta H= 66.8~{\rm{km/s/Mpc}},~
\Delta H = 10.5~{\rm{km/s/Mpc}},~
r_0=500~{\rm{Mpc}},~
\Omega_0 = 0.45~.
\end{equation}
The goodness of the fit is $\chi^2=172.6~ ({\chi^2}/{157}=1.10)$.
The confidence level contours with fixed values of $\Omega_0$ and
$H$ are shown in Fig. 1. For comparison with the homogeneous
case, the best fit nonflat $\Lambda$CDM has $\Omega_M = 0.5$,
$\Omega_\Lambda = 1.0$ with $\chi^2=175~({\chi^2}/{157}=1.11)$.
\begin{figure}[htp]
\begin{center}
\includegraphics[width=12.5cm]{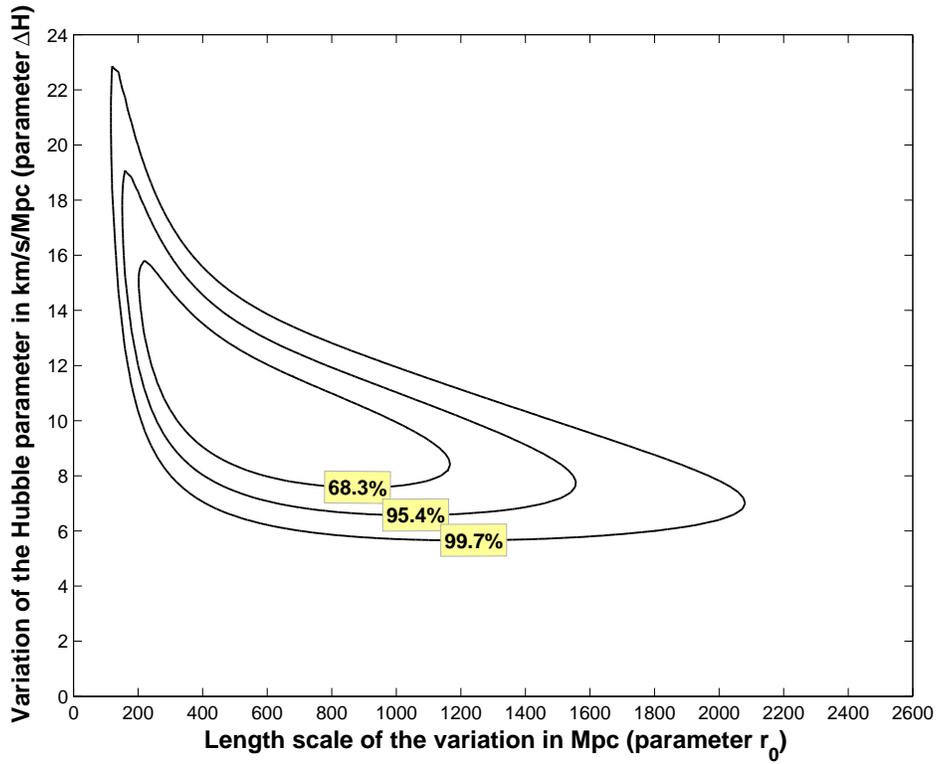}
\caption{Confidence level contours in the LTB model with
$\Omega_M(r) = {\rm{constant}} = 0.45$ and $H_0(r) =$ $56.3$
${\rm{km/s/Mpc}} + \Delta H e^{-r/r_0}$. From \cite{teppo}.}\label{figur1}
\end{center}
\end{figure}
What is perhaps surprising is the fact that the supernova fit is not only in qualitative agreement
with the observed homogeneity in galaxy surveys but also automatically yields a
value for the present-day matter density that is consistent with the observations.
The smallness ($\sim15 \%$) of the spatial variation in the
Hubble parameter is also somewhat surprising, considering that  it is of the same order as
the uncertainty of the
model-independent\footnote{Note that the smaller uncertainties
found in the CMB data analysis cannot be used here as those fits
assume that the entire universe is perturbatively close to the
homogeneous FRW model.} determination of the local Hubble rate
by the Hubble space telescope
\cite{Freedman:2000cf}. The variation of the Hubble parameter found
by Alnes, Amarzguioui and Gr{\o}n \cite{Alnes:2005rw}, who used
a different model Ansatz,
 has also
similar magnitude, but in contrast their model contains a large ($\sim 400 \%$)
variation in the matter density at scales larger than the current range of galaxy surveys.

\begin{figure}[h!]
\begin{center}
\includegraphics[width=12.5cm]{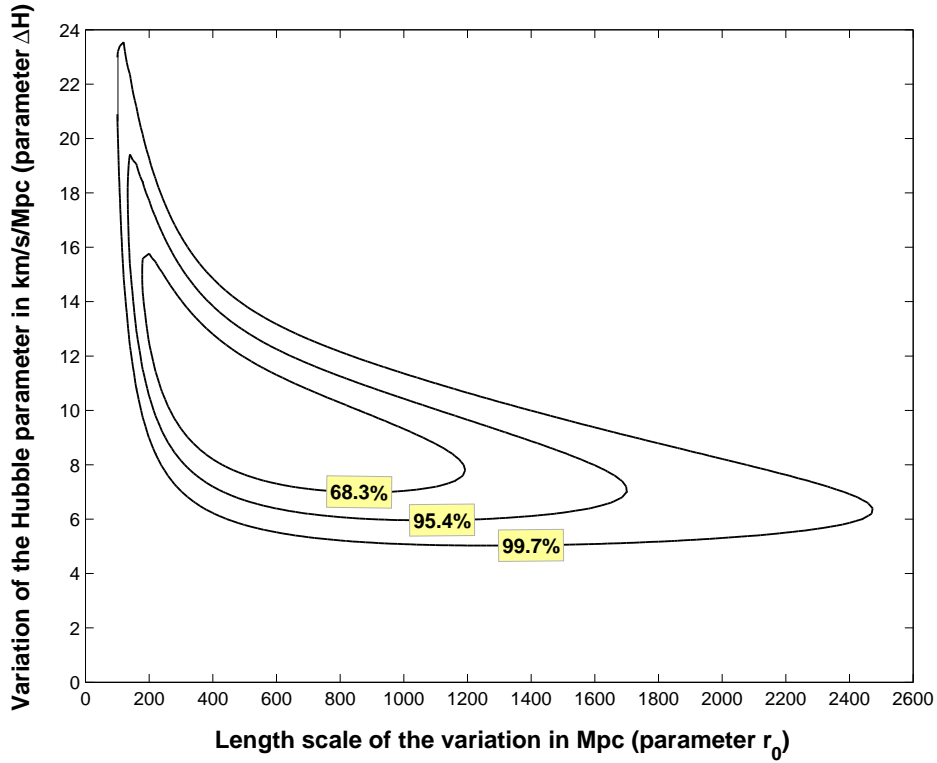}
\caption{Confidence level contours in the LTB model with perfectly
uniform present-day matter density: $H_0(r)=57$ ${\rm{km/s/Mpc}}+
\Delta H e^{-r/r_0}$, $\Omega_M(r) = 0.29\; (67$ ${\rm{km/s/Mpc}}
)^2/H_0^2(r)$. From \cite{teppo}.} \label{figur3}
\end{center}
\end{figure}

One can also fit data with Eq. (\ref{SNmodel}) together with a cosmological constant. Taking
$\Omega_M(r)+\Omega_\Lambda(r)=1$ one finds no improvement \cite{teppo}. If instead
of $\Omega_M =$ const. we assume a strictly uniform present-day
matter distribution with $\rho_M(r,t_0)={\rm{constant}}$,
which implies $H_0^2(r) \Omega_M(r) = {\rm{constant}}$, we may choose
the parametrization
\begin{eqnarray}
H_0(r) &=& H + \Delta H e^{-r/r_0}~,\nonumber\\
\Omega_M(r) &=& \Omega_0 (H + \Delta H)^2/(H + \Delta H
e^{-r/r_0})^2~.
\end{eqnarray}
The best fit values in this case are
\begin{equation}\label{SNfitlambda}
H+ \Delta H= 67~{\rm{km/s/Mpc}},~
\Delta H = 10~{\rm{km/s/Mpc}},~
r_0=450~{\rm{Mpc}},~
\Omega_0 = 0.29~.
\end{equation}
Here the goodness of the fit is $\chi^2=172.6$ $({\chi^2}/{157}=1.10)$.
The confidence level contours with $\Omega_0$ and $H$ fixed to their
best fit values are displayed in Fig. \ref{figur3}.

All these models have an inhomogenous Big Bang. One could also have
an inhomogenous expansion with a spatially constant age of the universe by choosing e.g.
\begin{eqnarray}\label{boundcond35}
H_0(r) &=& H \left[ \frac{\sqrt{1-\Omega_M(r)} - \Omega_M(r)
{\rm{arsinh}}\sqrt{\frac{1-\Omega_M(r)}{\Omega_M(r)}} }{
(1-\Omega_M(r))^{3/2}} \right] ~,\nonumber\\
\Omega_M(r) &=& \frac{\Omega_0}{(1+\delta e^{-r/r_0})^2}~.
\end{eqnarray}
The constraint of a simultaneous Big Bang leaves
us with only one free function. The best fit values are \cite{teppo}
\begin{equation}\label{SNfitsimBB}
H= 76.5~{\rm{km/s/Mpc}},~
\delta  = 1.21,~
r_0=1000~{\rm{Mpc}},~
\Omega_0 = 0.29~
\end{equation}
with
$\chi^2=175.5$ ($\chi^2/157=1.12$). Eq. (\ref{SNfitsimBB}) implies that the
Hubble function $H_0(r)$ varies from the value $H_0(0)=65$
${\rm{km/s/Mpc}}$ near us to its asymptotic value $H_0(r \gg
r_0)=52$ ${\rm{km/s/Mpc}}$. The age of the universe is then
$t_{{\rm{age}}} = 1/H = 12.8$ ${\rm{Gyr}}$. Similar
values have also been found in the model of ref. \cite{Alnes:2005rw}.

Thus simple and at least seemingly semirealistic LTB dust models can fit the
supernova data. The point to note is that although the LTB equations of motion do not in general permit
locally accelerated expansion, this does not exclude
the possibility that there can be an effective, volume averaged acceleration, where a scale
factor defined via the physical volume of some comoving region has a positive
double time derivative \cite{Alnes:2007}. However, it can be shown there is no effective average acceleration  \cite{teppo}
for the models considered above\footnote{Although fitting the supernova data does not require
accelerating expansion, for some profiles the LTB model may give rise to a suitably
defined average acceleration \cite{chuang}.
For a discussion on backreaction in LTB models,
see also \cite{syksy04}.}.

\section{Towards more realistic LTB models}\label{realistic}

Whether the supernova data combined with the CMB and LSS data would
nevertheless require an accelerating universe is an open
question; cosmological perturbation theory in LTB background is
still non-existent.

Some issues can be addressed, though. In particular,
when the LTB metric models a local underdensity, one may assume
that the evolution of perturbations is
identical to that in a homogeneous universe until the time of last
scattering. Adopting this approach, Alnes, Amarzguioui and Gr{\o}n \cite{Alnes:2005rw}
have considered in an approximation constraints arising from the position $l_1$ of first acoustic peak.
They find a shift relative to the concordance $\Lambda$CDM model that is given by
\begin{equation}\label{alnesshift}
\mathcal{S}=\frac{l_1}{l_1^{\Lambda CDM}}=0.01419(1-\phi_1)\frac{d_A}{r_s}~,
\end{equation}
where $d_A$ is the angular diameter distance to the last scattering surface,
given by Eq. (\ref{angdist}); this is the part that depends on the local underdensity, whereas
$r_s$, the sound horizon at recombination, and the (small) value of the parameter $\phi_1$
can be obtained from the conventional homogeneous model.
To be in agreement with the WMAP observations, the shift
parameter should be within the range $\mathcal{S} = 1.00 \pm 0.01$.
The locally underdense model depends on the density contrast parameter $\Delta\alpha$, functionally related
to $A(r,t)$ and  specifying
the difference between the two region,
the transition point $r_0$ from LTB to FRW, and the transition width $\Delta r/r_0$.
A set of parameter values that yields a good fit both to the supernova data
and the first CMB peak position can be found, as can be seen from Table~\ref{tab:model}.
%\squeezetable
\begin{table}[t]
\begin{tabular}{lcc}
\hline
Description & Symbol & Value\\
\hline
Density contrast parameter & $\Delta \alpha$ & 0.90 \\
Transition point & $r_0$ &  1.35 Gpc \\
Transition width & $\Delta r/r_0$ & 0.40 \\
Fit to supernovae & $\chi^2_{SN}$ & 176.5\\
Position of first CMB peak & $\mathcal{S}$ & 1.006 \\
Age of the universe & $t_0$ & $12.8 {\rm Gyr}$\\
Relative density inside underdensity & $\Omega_{m,in}$ & 0.20 \\
Relative density outside underdensity & $\Omega_{m,out}$ & 1.00 \\
Hubble parameter inside underdensity & $h_{in}$ & 0.65 \\
Hubble parameter outside underdensity & $h_{out}$ & 0.51 \\
Physical distance to last scattering surface & $D_{LSS}$ & 11.3 Gpc\\
Length scale of baryon oscillation from SDSS & $R_{0.35}$ & 107.1 \\
\hline
\end{tabular}
\caption{The best fit parameters of the locally underdense
  inhomogeneous model of \cite{Alnes:2005rw}.}
\label{tab:model}
\end{table}
Generically, for the void picture to work, one should have a local underdense region
that extends at least up to the nearby supernovae or about
300-400 Mpc/h.

These considerations hold if we occupy the exact center of the local
LTB universe. For an observer that is located off-center, the universe
appears to be anisotropic. Estimating the luminosity distance for an off-center observer is somewhat more
complicated task than in the case of an observer at the center  \cite{Alnes:2006pf, Biswas:2006ub}. One finds an anisotropic relation between
the redshifts and the luminosity distances of supernovae, which
however yields only a mild constraint as up to about 20 \% displacement from
the center is consistent with the data \cite{AlnesAmarzguioui}. In contrast, the constraint from the
CMB dipole appears to be very stringent, allowing only a displacement of
about 15 Mpc from the center of the underdense bubble \cite{Alnes:2006pf}.
This result is obtained by assuming that all of the observed dipole $a_{10}\sim 10^{-3}$ is due
to the displacement. A cancelation of the dipole due to our local peculiar motion towards the center of
the underdensity is a possibility that would allow for a larger displacement. Whether such a
peculiar motion can arise naturally or only by an accident, remains to be seen.

For an off-center observer the direction towards the center of the bubble singles out a special axis.
Therefore one could hope that a local LTB bubble could provide an explanation for the observed peculiar
alignments of the CMB quadrupoles and octopoles \cite{align}.
Because of the smallness of the displacement allowed by the dipole,
the quadru- and octopoles appear not to have enough power to explain their observed alignment \cite{Alnes:2006pf}, although
again the conclusion depends on the assumption that our local average motion has been accounted for  correctly.

Instead of a single underdensity, one could also consider an "onion model" with
a homogeneous background density, on top of which there are density fluctuations which are periodic as a
function of the radial coordinate. The observer sits in some generic position and looks at
sources along the radial direction, and the LTB dust solution incorporates the entire Universe.
To study this set-up, Biswas, Mansouri and Notari  \cite{Biswas:2006ub} have derived an expression for the luminosity
distance in an LTB metric for an off-centre observer.
The corrections due to underdensities to light propagation were found to have a tendency to cancel far away from the observer
because a radial light ray
unavoidably meets both underdense and overdense structures.
However, in the real universe light encounters hardly any
structure, so the cancellations might be an artifact of the onion model. Since in the real universe the photon
is mostly traversing voids it should get redshifted faster as the nonlinearities
increase with time and thereby effectively produce an apparent
acceleration. In the onion model
one can nevertheless mimic an accelerating
$\Lambda$CDM cosmology under certain special
conditions: the observer has to be located around a minimum of the density
contrast that is required to be quite high \cite{Biswas:2006ub}.

Yet another approach is the "Swiss cheese" model
of the inhomogeneous universe, where each spherical void is described by the LTB metric.
At the boundary of these regions the LTB metric is matched with the FRW metric that
describes the evolution between the inhomogeneities. One can then seek for the
modifications of the luminosity distance as the light passes through the underdense
regions \cite{tetradis}. In the extreme case where one assumes that light traverses the centers
of all the inhomogeneities along its path, assuming that the locations of the source and
 the observer are random and inhomogeneities have sizes of order 10 Mpc, the relative
 increase of the luminosity distance is however just of the order of a
 few percent near $z\simeq 1$. A qualitatively similar conclusion has been
 reached in \cite{biswasnotari}.

Structure formation and the smallness of CMB perturbations may in general pose a difficulty for
LTB models. For instance, for a class of inhomogeneities a homogeneous universe is actually a late
time attractor solution. This means that at earlier
times matter density and/or Hubble rate tends to be even more inhomogeneous than today. Whether
this presents an unsurmountable problem remains to be seen. Nevertheless,
the LTB model serves as a reminder that the interpretation of the cosmological data
is  not only quantitatively but even qualitatively very much model dependendent.
Therefore, all options should be carefully examined before firm conclusions can be drawn.
This is true in particular for dark energy, which is both an observational
and theoretical enigma.

\vspace{1.0cm}\noindent
{\bf Acknowledgements}
\\ \\
I thank Maria Ronkainen for help in preparing this
manuscript and Teppo Mattsson for many useful discussions on the LTB model.
This work is partially supported by the Academy of Finland grant
114419 and the Magnus Ehrnrooth foundation.

\end{document}